\documentclass[pdftex, twocolappendix, numberedappendix, iop, apj]{emulateapj}
\pdfoutput=1

\newcommand{\ud}{\,\mathrm{d}}

\newcommand{\nhat}{\boldsymbol{\hat n}}
\newcommand{\e}{e}
\newcommand{\lcdm}{$\Lambda$CDM}
\newcommand{\wmap}{\textit{WMAP}}
\newcommand{\planck}{\textit{Planck}}
\newcommand{\K}{\textit{K}}
\newcommand{\Ka}{\textit{Ka}}

\newcommand{\mpc}{\,\mathrm{Mpc}}

\newcommand{\ev}{\,\mathrm{eV}}

\newcommand{\ghz}{\,\mathrm{GHz}}

\newcommand{\uk}{\,\mu\mathrm K}

\newcommand{\ukarcmin}{\,\mu\mathrm{K\,arcmin}}

\usepackage{amsmath}
\usepackage{amssymb}
\usepackage{natbib}
\bibstyle{aa}

\usepackage{mathrsfs}
\usepackage{pgf}
\pgfrealjobname{fg}
\long\def\beginpgfgraphicnamed#1#2\endpgfgraphicnamed{\includegraphics{#1}}

\usepackage[utf8]{inputenc}
\usepackage[varg]{txfonts}

\usepackage[backref,breaklinks,colorlinks,citecolor=blue]{hyperref}        
\usepackage[all]{hypcap}

\righthead{Foreground Cleaned B-Modes on a Cut Sky}
\lefthead{Watts et al.}
\begin{document}

\title{Measuring the Largest Angular Scale CMB B-mode 
Polarization\\ with Galactic Foregrounds on a Cut Sky}

\author{Duncan J.~Watts\altaffilmark{1}}
\author{David Larson\altaffilmark{1}}
\author{Tobias A.~Marriage\altaffilmark{1}}
\author{Maximilian H.~Abitbol\altaffilmark{1,2}}
\author{John W.~Appel\altaffilmark{1}}
\author{Charles L.~Bennett\altaffilmark{1}}
\author{David T.~Chuss\altaffilmark{3,4}}
\author{Joseph R.~Eimer\altaffilmark{1}}
\author{Thomas Essinger-Hileman\altaffilmark{1}}
\author{Nathan J.~Miller\altaffilmark{4}}
\author{Karwan Rostem\altaffilmark{1,4}}
\author{Edward J.~Wollack\altaffilmark{4}}
\altaffiltext{1}{Department of Physics and Astronomy, Johns Hopkins University\\
3701 San Martin Drive, Baltimore, Maryland, United States}
\altaffiltext{2}{Department of Physics, Columbia University\\
538 West 120th Street, New York, New York, United States}
\altaffiltext{3}{Department of Physics, Villanova University \\
800 E.~Lancaster Ave., Villanova, Pennsylvania, United States}
\altaffiltext{4}{Observational Cosmology Laboratory, Code 665, NASA,\\
Goddard Space Flight Center, Greenbelt, Maryland, United States}

\email{dwatts@jhu.edu}

\begin{abstract}
We consider the effectiveness of foreground cleaning in the recovery of Cosmic
Microwave Background (CMB) polarization  sourced by gravitational waves for
tensor-to-scalar ratios in the range $0<r<0.1$.  Using the planned survey area,
frequency bands, and sensitivity of the Cosmology Large Angular Scale Surveyor
(CLASS), we simulate maps of Stokes $Q$ and $U$ parameters at 40, 90, 150, and
220 GHz, including realistic models of the CMB, diffuse Galactic thermal dust
and synchrotron foregrounds, and Gaussian white noise.  We use linear
combinations (LCs) of the simulated multifrequency data to obtain  maximum
likelihood estimates of $r$, the relative scalar amplitude $s$, and LC
coefficients.  We find that for 10,000 simulations of a CLASS-like experiment
using only measurements of the reionization peak ($\ell\leqslant23$), there is a 95\%
C.L.~upper limit of $r<0.017$ in the case of no primordial gravitational waves.
For simulations with $r=0.01$, we recover at 68\%
C.L.~$r=0.012^{+0.011}_{-0.006}$.  The reionization peak corresponds to a
fraction of the multipole moments probed by CLASS, and simulations including
$30\leqslant\ell\leqslant100$ further improve our upper limits to $r<0.008$ at 95\%
C.L.~($r=0.010^{+0.004}_{-0.004}$ for primordial gravitational waves with
$r=0.01$).  In addition to decreasing the current upper bound on $r$ by an order
of magnitude, these foreground-cleaned low multipole data will achieve a cosmic
variance limited measurement of the E-mode polarization's reionization peak.
\end{abstract}


\section{Introduction}

All astronomical data on cosmological scales conform  to a six-parameter model
of the Universe (\lcdm{}) with dark matter and a cosmological constant as the
dominant components \citep{wmap9, planck}.  The inflationary paradigm, which
postulates a short period of exponential expansion in the early Universe,
accounts for features in \lcdm{}, including the high degree of homogeneity and
flatness and the scalar spectral index $n_s\lesssim1$
\citep[e.g.][]{guth, starobinsky, kazanas, mukhanov, einhorn, linde, albrecht,
wmap9, wmapfinal}.  One of inflation's predictions is a super-horizon stochastic
gravitational wave background that induces polarization in the CMB
\citep{polnarev}.  We can decompose the CMB's polarization field into E-modes
with $(-1)^\ell$ parity, and B-modes with $(-1)^{\ell+1}$ parity
\citep{seljzald, kks, polprimer}.  An inflationary B-mode signal can only come
from primordial gravitational waves (tensor fluctuations of the metric), so a
measurement of such a signal would be strong evidence for an inflationary epoch.
We quantify constraints on B-modes in terms of the ratio of tensor fluctuations
to scalar (density) fluctuations, $r$, evaluated at $k=0.05\mpc^{-1}$.  The
current upper limit on tensor fluctuations ($r<0.09$) comes from a combination
of \planck{} and BICEP2 measurements \citep{bkp, planckinflation}.

\begin{figure*}
	\centering
        \resizebox{0.7\textwidth}{!}{
		\beginpgfgraphicnamed{f1}
        \input{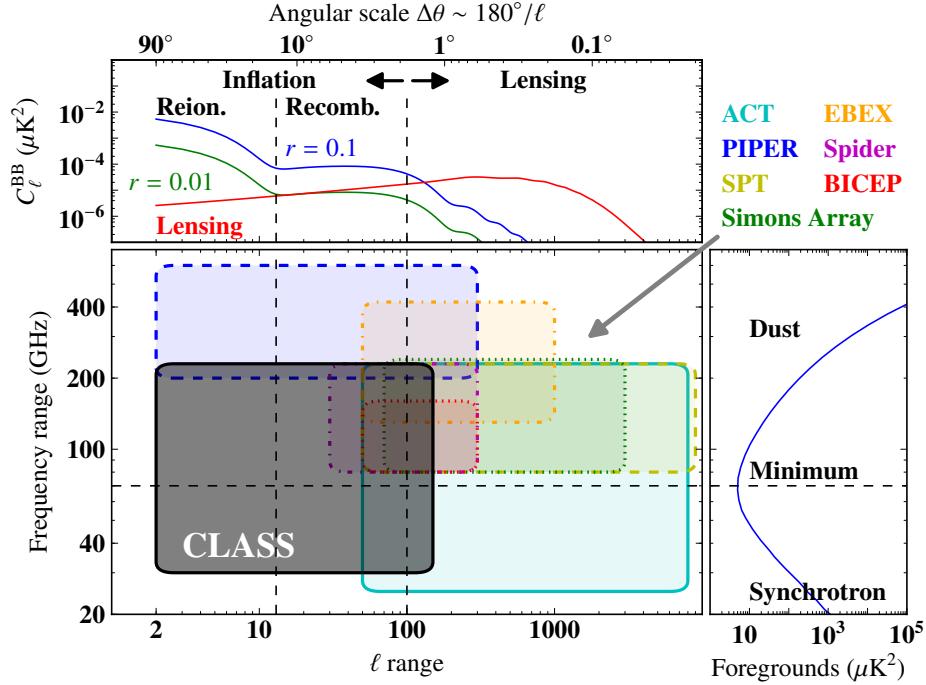}
        \endpgfgraphicnamed
        }
        \caption{
	We display schematically the regions of purview corresponding to
	balloon-borne and ground-based CMB polarization experiments.  CLASS is
	unique in measuring both the reionization and recombination peaks while
	straddling the foreground minimum.  As upper limits on $r$ decrease,
	inflationary B-modes will dominate over lensing at increasing larger
	scales.  The foreground model shown comes from \citet{planckfg} using
	measurements of 73\% of the sky. We plot the frequency dependence of
	foregrounds in thermodynamic temperature units.  }
        \label{fig:compare}
\end{figure*}

The Cosmology Large Angular Scale Surveyor (CLASS) is a ground-based experiment
that will observe 70\% of the sky from Cerro Toco in the  Atacama Desert at
frequencies of 40, 90, 150, and 220 GHz \citep{class, class_spie,
classdetectors,classpolarimeters}. First light for the experiment is imminent.
CLASS's $90\ghz$ band has a projected sensitivity of $10\ukarcmin$, an
improvement over the \planck{} $100\ghz$ map \citep[$118\ukarcmin$ at
high-$\ell$,][] {planckoverview2015}, and will achieve the measurement stability
required to reach low multipoles using front end modulation by a variable-delay
polarization modulator (VPM) \citep{vpm}.  CLASS is currently the only planned
sub-orbital mission exploring the combination of frequency and multipole ranges
described above.  CLASS will probe the reionization peak in the BB power
spectrum ($\theta\gtrsim2^\circ$) along with other missions including PIPER
\citep{piper}, QUIJOTE \citep{quijote}, LSPE \citep{lspe}, and GroundBird
\citep{groundbird}.  CLASS will also probe the recombination peak explored by
BICEP2 \citep{bicep2}, SPTpol \citep{sptpol}, ABS \citep{abs}, ACTPol
\citep{actpol}, POLARBEAR \citep{polarbear}, EBEX \citep{ebex}, and SPIDER
\citep{spider}.  In contrast with other surveys that focus on higher multipoles
or have different frequency ranges, CLASS probes a unique combination of
frequency and multipole space, as illustrated in Figure \ref{fig:compare}.  In
addition to constraining inflation at all of the angular scales where B-modes
are predominantly inflationary, we show that a noise-dominated BB spectrum
$C_\ell\propto\ell^{-2}$ has signal-to-noise that scales as $\ell^{-3/2}$, as
opposed to the well-known result $\ell^{1/2}$ found in the cosmic variance
limit.  In this era of initial measurements of B-mode polarization, we gain more
information from large angular scale measurements than conventional wisdom would
suggest, as we show in \S\ref{sec:appendix_a}.

In this paper we explore the recovery of B-modes in the presence of foregrounds,
instrument noise, and a cut sky. This analysis will focus on a subset of the
simulated CLASS data using only the multipoles $\ell\leqslant23$, to distill the
information available from measurements of the reionization peak alone. In
\S\ref{sec:simulations} we describe simulations of the CMB with $0<r<0.1$ and
foreground components in 40, 90, 150, and 220 GHz frequency bands using the
CLASS sensitivity and sky coverage. In \S\ref{sec:likelihood} we describe the
maximum likelihood method for recovering the primordial B-mode signal in the
presence of foregrounds. We forecast constraints on $r$ from this method applied
to the simulations in \S\ref{sec:recovery}.

\capstartfalse
\begin{deluxetable}{cccccccc}
\tabletypesize{\footnotesize}
\tablecolumns{6} 
\tablewidth{0pt} 
\tablecaption{Noise Contribution from Rescaled Foreground Templates}
\tablehead{
	\colhead{Frequency $\nu$}					&
	\colhead{$w_{\nu}^{-1/2}$} 					&
	\colhead{$\alpha_\mathrm S^\nu$\tablenotemark{$\dagger$}}	&
	\colhead{$\alpha_\mathrm S^\nu w_{40}^{-1/2}$}			&
	\colhead{$\alpha_\mathrm D^\nu$\tablenotemark{$\dagger$}}	&
	\colhead{$\alpha_\mathrm D^\nu w_{220}^{-1/2}$}
	\\
	\colhead{(GHz)}							&
	\colhead{($\mu$K\,arcmin)} 					&
	\colhead{}							&
	\colhead{($\mu$K\,arcmin)} 					&
	\colhead{}							&
	\colhead{($\mu$K\,arcmin)}					
}
\startdata
         40 	& 39 	& 1	& 39 	& 0.022	& 0.95
         \\
         90 	& 10 	& 0.103	& 4.02 	& 0.095	& 4.09
         \\
         150 	& 15 	& 0.032	& 1.25 	& 0.306	& 13.2
         \\
         220 	& 43 	& 0.018	& 0.70 	& 1	& 43
\enddata
\tablenotetext{$\dagger$}
{Values for the amplitude of synchrotron and dust templates at frequency $\nu$,
	$\alpha_i^\nu$, come from Equation \ref{eq:alpha} and assume
$\beta_\mathrm S=-3$ and $\beta_\mathrm D=1.6$.}
\tablecomments{
For each band we list the expected 5-year polarized noise at each frequency
($w_{\nu}^{-1/2}$).  The columns for $\alpha_i^\nu w_{\nu_i}^{-1/2}$ show the
expected noise contribution in each band when we rescale the 40 and 220 GHz
channels according to the synchrotron and dust template scalings used in
simulations (Equation \ref{eq:alpha}).  The values of $w_{\nu}^{-1/2}$ are
estimated in \cite{class_spie}.  } \label{tbl:noises} \end{deluxetable}
\capstarttrue

\section{Multifrequency Simulations}
\label{sec:simulations}

We use publicly available data to simulate the polarized Galactic synchrotron
and dust foregrounds.  Our synchrotron map templates are based on the \wmap{}9
\K{}-band ($23\ghz$) polarization data \citep{wmapfinal}.  Synchrotron
polarization dominates the data in this band with negligible contribution from
CMB polarization.    Our dust polarization templates are based on the \planck{}
$353\ghz$ maps \citep{planckoverview2015}.   In units of antenna temperature
$T_A(\nu)\equiv c^2I_\nu/2k_\mathrm B\nu^2$, with $I_\nu$ the specific
intensity, we approximate the modified blackbody emission of dust as a power law
with index $\beta_\mathrm D=1.6$ \citep{polfreq} and the synchrotron emission as
a power law with $\beta_\mathrm S=-3.0$ \citep{wmapfinal, specvar}.  Using these
power law indices we rescale the \wmap{} and \planck{} data to create templates
for synchrotron and dust at 40 and 220 GHz, respectively.  The minimum
contamination from foregrounds is around 70 GHz \citep{wmapfinal,planckfg},
which, along with the location of atmospheric water and oxygen lines, informs
the CLASS experiment's choice of band center frequencies \citep{class}.

We estimate CLASS's sensitivity to tensor modes using random realizations of
Gaussian CMB fluctuations with white Gaussian detector noise, all with the same
model of foreground contamination.  We use the fiducial \lcdm{} parameters from
\wmap{}9 \citep{wmap9}  to simulate the scalar perturbation field, with a tensor
contribution added in via
$C_\ell=rC_\ell^\mathrm{tensor}+sC_\ell^\mathrm{scalar}$. The contribution from
gravitational lensing is included in the $C_\ell^\mathrm{scalar}$ term.
We generate the
theoretical spectra $C_\ell$ using CAMB,\footnote{\url{http://camb.info}}
smoothed in multipole space using a Gaussian window function with
$\theta_\mathrm{FWHM}=15^\circ$ to reduce aliasing from pixelization.  We derive
random $Q$ and $U$ maps from the theoretical $C_\ell$ values using the Healpix
package \texttt{synfast} \citep{healpix} with pixels at resolution
$\theta_\mathrm{pix}\sim7^\circ$ ($N_\mathrm{side}=8$),\footnote{Healpix maps
	have resolutions denoted $r=0,1,2,\ldots$. Each of the 12 lowest
	resolution pixels that characterize $N_\mathrm{side}=1$ divides into
	$N_\mathrm{side}\times N_\mathrm{side}$ regions where
	$N_\mathrm{side}\equiv 2^r$, and the total number of pixels is
	$N_\mathrm{pix}=12N_\mathrm{side}^2$ with characteristic pixel size
	$\theta_\mathrm{pix}\sim58.6^\circ/N_\mathrm{side}$. The full
	documentation is at \url{http://healpix.sf.net}.
}
including multipoles $2\leqslant\ell\leqslant23$.  We then add dust and synchrotron
polarization to each band, along with white Gaussian noise corresponding to the
experiment sensitivity, given here in units of $\mu$K\,arcmin (see Table
\ref{tbl:noises}).

Given CLASS's wide scan strategy from its site in the Atacama Desert at latitude
$-23^\circ$ \citep{class_spie}, we include data with declination
$-73^\circ<\delta<27^\circ$ so that $f_\mathrm{sky}\simeq0.7$. We mask out the
Galactic equator based on the \wmap{} P06 mask, defined in \cite{pol_mask},
resulting in a final $f_\mathrm{sky}\simeq0.5$.

\begin{figure}
        \resizebox{0.43\textwidth}{!}{
		\beginpgfgraphicnamed{f2}
        \input{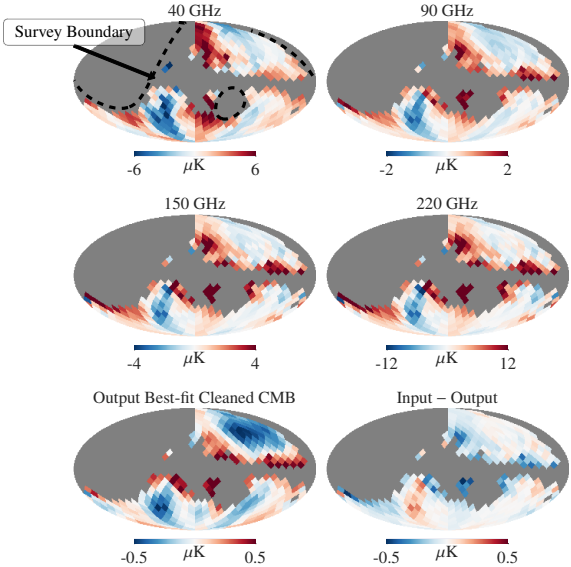}
        \endpgfgraphicnamed
        }
	\caption{ The top four plots show simulated CLASS observations of $Q$
		polarization assuming spatially varying spectral indices (as
		defined in Equation \ref{eq:alpha}).  The bottom two plots show
		(left) the output maximum likelihood CMB map with the simulation's best-fit
		LC parameters as described in \S\ref{sec:likelihood} and (right)
		the difference between the input CMB and the output maximum
		likelihood estimate.  The difference map shows the effectiveness
		of our method for reconstructing the CMB, and the temperature
		range is representative of the noise in our map. Note
		that each of these maps uses a different color scale.  We
		present the maps in Galactic coordinates (Mollweide projection)
		with gray pixels representing regions of the sky outside of the
		CLASS survey boundary or inside the Galactic mask.  The black
		dashed lines on the 40 GHz map denote the declination limits of
	$-73^\circ$ and $27^\circ$ in celestial coordinates.  }
	\label{fig:cmb_dif}
\end{figure}

We define the polarization maps $\boldsymbol P$ as a vector with the
$Q$ and $U$ Stokes parameters as components.
We use $\boldsymbol P^\mathrm S$ and $\boldsymbol
P^\mathrm D$ as the synchrotron and dust templates at $40\ghz$ and $220\ghz$
respectively, and $\boldsymbol P^\mathrm{CMB}$ the simulated CMB. This allows us
to generate the simulated multifrequency CLASS polarization data,
\begin{equation}
\label{eq:sims}
\boldsymbol P^\nu=\alpha_\mathrm S^\nu\boldsymbol P^\mathrm S+\alpha_\mathrm
D^\nu\boldsymbol P^\mathrm D+\boldsymbol P^\mathrm{CMB}+\boldsymbol
N^\nu
\end{equation}
where $\boldsymbol N^\nu$ contains the  Gaussian white noise maps for $Q$ and
$U$, with the amplitude of the noise determined by the survey sensitivity given
in Table \ref{tbl:noises} \citep{class_spie}.  Because we define the power law
spectral indices $\beta_i$ in terms of antenna temperature and our data are in
units of thermodynamic temperature, we scale the foreground channels using the
factor
\begin{equation}
\alpha_i^\nu(\nhat)\equiv\left(\frac\nu{\nu_i}\right)^{\beta_i(\nhat)}
\frac{g(\nu)}{g(\nu_i)}
\label{eq:alpha}
\end{equation}
with $g(\nu)\equiv\partial T/\partial T_A=(\e^x-1)^2/(x^2\e^x)$ the conversion
factor from thermodynamic to antenna temperatures with $x\equiv
h\nu/kT_\mathrm{CMB}=\nu/56.78\ghz$. These $\alpha_i^\nu$ give the relative
strength of each foreground component $i$ normalized by its strength at template
frequencies $\nu_i$. As described above these templates are based on the \wmap{}
\K-band for synchrotron and \planck{} 353 GHz band for dust scaled to
$\nu_\mathrm S=40\ghz$ and $\nu_\mathrm D=220\ghz$, respectively.  For these
chosen frequencies and spectral indices, these strengths are $\alpha_\mathrm
S^{90}=0.103$ and $\alpha_\mathrm D^{90}=0.095$ at $90\ghz$ while for $150\ghz$,
we have $\alpha_\mathrm S^{150}=0.032$ and $\alpha_\mathrm D^{150}=0.305$, as
shown in Table \ref{tbl:noises}.  We display sample maps of our simulated CLASS
data in Figure \ref{fig:cmb_dif}, along with the best-fit cleaned data and the
residuals, to be described in \S\ref{sec:likelihood}.

The spatial dependence of the spectral indices is based on the pre-flight
\planck{} Sky Model (PSM) \citep{psm}, which we display in Figure
\ref{fig:regions}. As we demonstrate below, the PSM is 
consistent with current data. The $\beta_i$ from this model come from intensity
measurements of foreground components, and are not expected to be the same as
the polarized power law indices.  For this work, we rescale the PSM spectral
indices to have mean values $\langle\beta_\mathrm S\rangle=-3.0$ and
$\langle\beta_\mathrm D\rangle=1.6$.
The amplitude of synchrotron spectral index
fluctuations in the \wmap{} estimate derived from the \K{} and \Ka{} band data
\citep{wmapfinal, specvar} is broadly consistent with the variation in the PSM's
$\beta_\mathrm S$ \cite[model 4]{psm_synch}.  The variation in the dust spectral
index has variation $\Delta\beta_\mathrm D$ \citep{polfreq} that is consistent
with a constant spectral index (see \S\ref{sec:appendix_b}) and hence with the
small variation in the PSM \citep[model 8]{fds}. 
Note that we only use the pre-flight PSM for
spectral variation, while we derive the 40 and 220 GHz polarization templates
from
\wmap{} and \planck{} data respectively, as described above. However, the CLASS
dataset itself will improve on this and be used in our final analysis.

In this study we focus on the effects of foregrounds and a cut sky on the
recoverability of primordial B-modes. To isolate these effects we do not include
potential instrumental systematics. For measurements that use a  VPM to suppress
$1/f$ noise, \citet{miller} find that B-mode power from systematic effects can
be suppressed to levels below the primordial level for $r=0.01$, which suggests
such effects are negligible in the context of this work.

\section{Pixelized Likelihood Analysis} \label{sec:likelihood}

\begin{figure} 
    \resizebox{0.43\textwidth}{!}{ 
        \beginpgfgraphicnamed{f3}
        \input{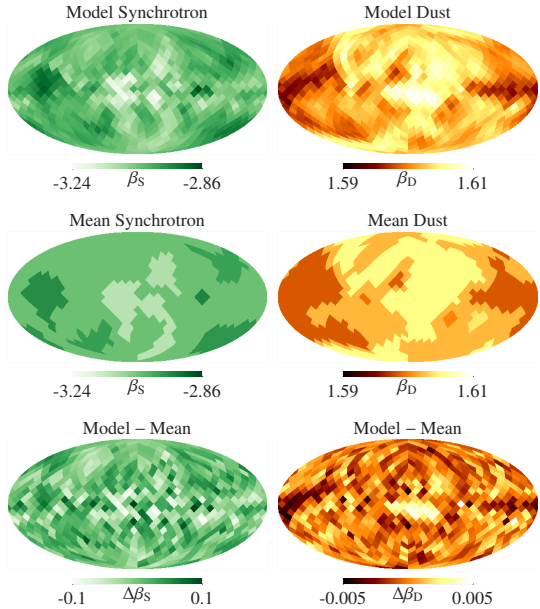} 
        \endpgfgraphicnamed
        } 
        \caption{ The top row of
        foreground spectral indices from the PSM \citep{psm} are used in our
        simulations. We model the variation in spectral index by breaking the
        sky into 14 regions (middle row, labelled ``Mean'') with distinct LC fit
        coefficients.  The $\beta_i$ in the middle row are the means of the PSM
        spectral indices within each region.  This modelling trades off accuracy
    at small scales for computational efficiency.  } \label{fig:regions}
\end{figure}

\begin{figure*} 
    \begin{center} 
        \resizebox{0.84\textwidth}{!}{
            \beginpgfgraphicnamed{f4} 
            \input{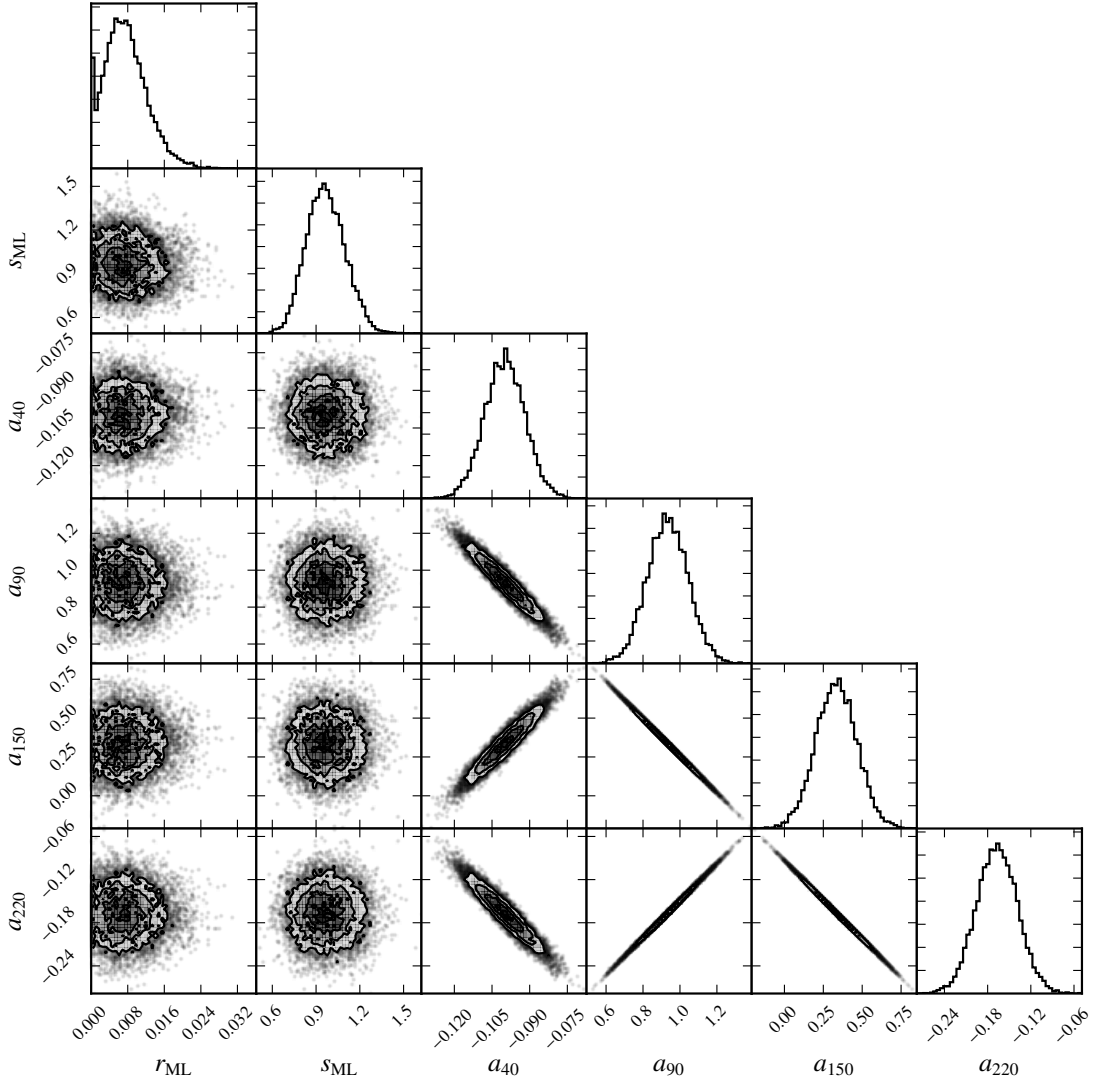} 
        \endpgfgraphicnamed 
    }
    \end{center} 
    \caption{Distribution of the output maximum likelihood
        parameters for $10^4$ simulations using the sensitivities from Table
        \ref{tbl:noises} with an input $r=0.01$.  The simulation features
        spatially varying spectral indices and the distribution of LC
        coefficients $a_\nu$ are displayed for one of the fourteen regions
        in Figure \ref{fig:regions}.  This demonstrates that the maximum
        likelihood solutions for $r_\mathrm{ML}$ and $s_\mathrm{ML}$ are
        uncorrelated with each other and the LC coefficients.  }
        \label{fig:corr} 
\end{figure*}

We base our analysis on methods developed in \cite{efstathiou} and
\cite{katayama}, and extend the Gaussian likelihood formalism to an arbitrary
number of frequency channels, each with its own linear fit coefficient. We
define the likelihood \begin{equation} \mathcal
    L\propto\frac{\exp[-\frac12\boldsymbol x(a_\nu)^T\boldsymbol
    C_p^{-1}(r,s,a_\nu)\boldsymbol x(a_\nu)]}{\sqrt{\det[\boldsymbol
    C_p(r,s,a_\nu)]}}, \label{eq:like} \end{equation} where \begin{equation}
    \label{eq:cmb_est} \boldsymbol x=\sum_\nu^n a_\nu\boldsymbol P^{\nu},\quad
    \sum_\nu^n a_\nu=1 \end{equation} is the CMB map cleaned using $n$ frequency
channels, with $\sum_\nu a_\nu=1$ preserving the CMB's frequency spectrum.  We
achieve this constraint defining $a_{90}$ as $1-a_{40}-a_{150}-a_{220}$.  We
define the pixel-space covariance matrix,
\begin{equation} 
    \boldsymbol C_p = r\boldsymbol C_p^\mathrm{tensor}+ s\boldsymbol C_p^\mathrm{scalar}
    +\boldsymbol I_p\sigma^2.
    \label{eq:cov} 
\end{equation}
Here $\boldsymbol I_p$ is the identity matrix in pixel space, and
$\sigma^2=\sum_\nu(a_\nu\sigma_\nu)^2$ is the survey noise.  The  $\boldsymbol
C_p$'s are the theoretical signal covariance matrices, which themselves come
from the power spectra, $C_\ell^\mathrm{BB}$ and $C_\ell^\mathrm{EE}$, assuming
uncorrelated E- and B-modes, as detailed in Appendix A of \cite{katayama}.  We
vary $s$, the amplitude of scalar fluctuations normalized by a fiducial value,
to mitigate spurious correlations between the CMB and foregrounds in the maximum
likelihood fitting as in \S5.1 of \cite{katayama}.  We use uniform unbounded
priors on all parameters except for the tensor-to-scalar ratio and the 40 GHz
and 220 GHz coefficients. We impose the prior $r\geqslant0$ to avoid singular
covariance matrices in Equation \ref{eq:cov}, and we impose the prior that the
40 and 220 GHz coefficients are negative to avoid numerical errors (which does
not affect our overall results, as seen in Figure \ref{fig:corr}).  The product
of our likelihood and priors is proportional to the Bayesian posterior
probability distribution for our parameters.  We emphasize that this method
differs from the internal linear combination method described in
\cite{efstathiou_lowell} due to our simultaneous variation of cosmological and
foreground LC parameters and also due to the presence of the $\boldsymbol
C_p^{-1}$ weighting for the solution of LC coefficients.  When we maximize the
likelihood (Equation \ref{eq:like}) using the covariance matrix (Equation
\ref{eq:cov}), the maximum likelihood LC coefficients, $\{a_\nu\}$, we create a
map, $\boldsymbol x$, that best approximates a realization of the CMB with
Gaussian noise. The modified foreground removal method is effective because the
pixel-pixel correlations, $\boldsymbol C_p$, drive the fit so that the mean of
the map, $\langle\boldsymbol x\rangle$, is zero, with variance corresponding to
CMB fluctuations and instrumental white noise.

We obtain an estimate of the CMB, $\boldsymbol x$, by finding the parameters
$r$, $s$, $a_\mathrm{40}$, $a_\mathrm{90}$, $a_\mathrm {150}$, and
$a_\mathrm{220}$ that maximize the likelihood $\mathcal L$ with the priors
$r\geqslant0$ and $a_{40},a_{220}\leqslant0$.  We find that although a Monte Carlo Markov
Chain (MCMC) estimation gives posterior distributions and suitable estimators
for these parameters, using the \texttt{scipy} minimization routine
\texttt{fmin\_l\_bfgs\_b}, an implementation of the limited-memory
Broyden-Fletcher-Goldfarb-Shanno (BFGS) algorithm \citep{bfgs1}, gives estimates
of the maximum likelihood values for parameters
$\{r_\mathrm{ML},s_\mathrm{ML},a_{\nu,\mathrm{ML}}\}$ at a fraction of the time
required to compute the full MCMC posterior distributions.  We plan to use MCMC
methods for the analysis of the actual multi-frequency CLASS observed maps, but
cosmic variance can cause different posterior distributions for the same
underlying value of $r$.  For our $\sim10^4$ simulations with random
realizations of CMB and noise, the distribution of maximum likelihood values for
$r$ is comparable to the width of the posterior distribution of a single Monte
Carlo chain.  For forecasting purposes, we therefore run large numbers of
simulations and study the posterior distribution of maximum likelihood values of
the model parameters.

We also model spatially varying spectral behavior in the foregrounds.  In
practice, we split the map into fourteen separate foreground regions defined by
the intersection of unmasked dust and synchrotron regions shown in Figure
\ref{fig:regions}, which results in a 44-parameter fit (four $a_\nu$
coefficients with one constraint $\times$ 14 regions, $s$, and $r$) for varying
spectral behavior.  We define the regions in the middle row of Figure
\ref{fig:regions} using the PSM components in the top row.  To construct these
regions, we smooth the PSM $\beta_i$ maps, find all local minima and maxima, and
choose contiguous pixels with roughly constant spectral index values.  We choose
thresholds such that the regions are small enough to make a constant index a
good approximation, and large enough that we maintain computational efficiency
that follows from the smaller number of parameters.  Defining these regions
according to existing measurements of spectral variation is the only part of our
analysis that requires data external to CLASS. When the CLASS data are
available, they will be the best probe of this large-scale position dependence
of the foreground spectral indices.

We discuss the distribution of maximum likelihood $r$ values in the following
section.  For typical simulations with input $r=0.01$, the other parameters for
a single representative region have sample mean and standard deviation
$s=0.97\pm0.13$, $a_{40}=-0.1\pm0.01$, $a_{90}=0.9\pm0.1$, $a_{150}=0.3\pm0.1$,
and $a_{220}=-0.16\pm0.03$. Both $a_{40}$ and $a_{220}$ are negative because
data at these frequencies serve to remove foregrounds, while the 90 and 150 GHz
coefficients contribute positively, with most of the weight coming from
$a_{90}$.  We display the distribution of ML values for these parameters and $r$
in Figure \ref{fig:corr}.

\section{Recovery of $\lowercase{r}$} \label{sec:recovery}

If the uncertainty were only dependent on instrument noise and not a function of
the cosmological value of $r$, we would choose a threshold based on this
uncertainty, such as $3\sigma_r$, and claim this as a detection limit. However,
the prediction problem is more complicated, as $\sigma_r$ depends on $r$ as well
as dust and synchrotron emission in the Galaxy through chance
cross-correlations.  We focus on the effect of varying $r$ on the detection
limit, which we will not know until real measurements are made.  We find the
distribution of recovered maximum likelihood values $r_\mathrm{ML}$ for a given
input $r$.

\begin{figure} 
    \begin{center} 
        \beginpgfgraphicnamed{f5} 
        \input{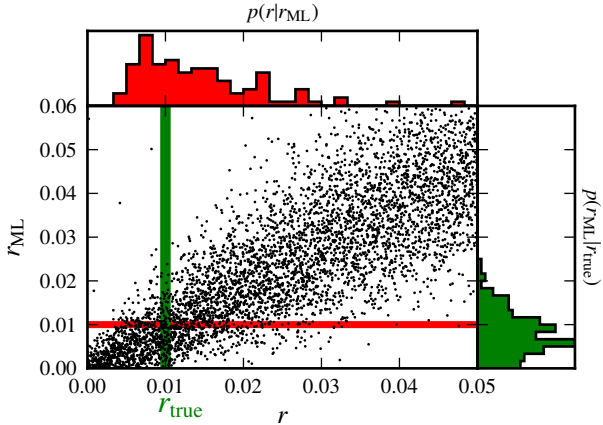}
        \endpgfgraphicnamed 
    \end{center} 
    \caption{ Each point in the bottom-left
    plot represents a unique realization of the simulations described in
    \S\ref{sec:simulations}.  Given an input tensor-to-scalar ratio $r$ we
    compute the distribution of recovered maximum likelihood values
    $r_\mathrm{ML}$.  We also display sample distributions of $r_\mathrm{ML}$
    given an $r_\mathrm{true}=0.01\pm0.0005$ (green histogram) in the right
    plot, and of $r$ given an $r_\mathrm{ML}=0.01\pm0.0005$ (red histogram) in
    the top plot. To obtain our posterior distribution of $r$ given a value
    $r_\mathrm{true}$, we integrate the product of these distributions over
    $r_\mathrm{ML}$ (Equation \ref{eq:bayes}).  } \label{fig:scatter}
\end{figure}

Using the simulations described in \S\ref{sec:simulations} and the maximum
likelihood method described in \S\ref{sec:likelihood}, we generate realizations
of CLASS-like data and infer best-fit parameters from them. The maps all have
scalar fluctuations with a fixed normalization ($s=1$) and the same foreground
simulation from \S\ref{sec:simulations},  while the tensor modes have varying
strength $0<r<0.1$ uniformly sampled with one $r$ every $\delta
r=5\times10^{-6}$. We display a subset of these simulations in Figure
\ref{fig:scatter}, marginalized over $s$ and the LC coefficients.

Actual CLASS data will provide a maximum likelihood estimate $r_\mathrm{ML}$.
Using simulated data similar to those from Figure \ref{fig:scatter}, we will
estimate the distribution of $r$ using the slice $p(r|r_\mathrm{ML})$.  Without
knowledge of $r_\mathrm{ML}$, however, we need to marginalize over the possible
values of $r_\mathrm{ML}$ that result from cosmic variance given a ``true'' $r$.

We estimate the constraining power of CLASS prior to a specific measurement
$r_\mathrm{ML}$.  Our goal is to calculate a probability distribution
$p(r|r_\mathrm{true})$ that indicates our best estimate of $r$ given some true
value, $r_\mathrm{true}$. From simulations, we calculate the probability of a
specific value of $r$ given a measured $r_\mathrm{ML}$, $p(r|r_\mathrm{ML})$,
and the probability of obtaining a given maximum likelihood result
$r_\mathrm{ML}$ given the true value of $r$, $p(r_\mathrm{ML}|r_\mathrm{true})$.
Using this information, we can marginalize over the nuisance parameter,
$r_\mathrm{ML}$, i.e.  
\begin{equation} p(r|r_\mathrm{true})=\int
    p(r|r_\mathrm{ML}) p(r_\mathrm{ML}|r_\mathrm{true})\ud r_\mathrm{ML}.
    \label{eq:bayes} 
\end{equation}
The quantity $p(r|r_\mathrm{true})$ 
represents the average of a CLASS-like experiment's 
posterior distribution for $r$ over many realizations, given that the true 
amplitude of tensor fluctuations is $r_\mathrm{true}$.
This statistic is motivated by noting a single CLASS-like experiment will
produce a probability distribution $p(r|r_\mathrm{ML})$. The random variable
$r_\mathrm{ML}$ is a draw from the distribution
$p(r_\mathrm{ML}|r_\mathrm{true})$, and is not of inherent interest to this
discussion. The $p(r_\mathrm{ML}|r_\mathrm{true})$ function acts as a kernel for
a weighted average, and is used here to create a weighted average of each
$p(r|r_\mathrm{ML})$, which results in the distribution $p(r|r_\mathrm{true})$.
We illustrate this process in Figure
\ref{fig:scatter} in which the histogram across the top gives a single
realization of $p(r|r_\mathrm{ML})$ and the histogram on the right shows
$p(r_\mathrm{ML}|r_\mathrm{true})$ for a specific value $r=r_\mathrm{true}$ for
the input model.  In this formalism, we can obtain a distribution of expected
estimates of $r$ given $r_\mathrm{true}$. 

We also  estimate the constraining power of CLASS's high-multipole data, with
$\theta_\mathrm{pix}\sim27.5'$ ($N_\mathrm{side}=64$, $\ell\lesssim100$).  The
model used here has a fixed scalar fluctuation amplitude $s=1$ and spatially
constant spectral indices. These simplifications are justified because
foreground power spectra decay approximately as $C_\ell\propto\ell^{-2.5}$, and
spurious correlations between the CMB and foregrounds are negligible at these
scales \citep{pol_mask,planckdustpol}.  We use an internal linear combination
(ILC) method similar to that used in the \wmap{} analysis \citep{wmap_ilc} to
clean foregrounds, and then use PolSpice \citep{polspice} to extract the B-mode
power spectrum for multipoles $30\leqslant\ell\leqslant100$.  By simulating maps with
varying $r$ as in the low-$\ell$ simulations, we find that we can describe the
likelihood of $r$ from the high-$\ell$ distribution with a Gaussian probability
distribution $p(r|r_\mathrm{true})\propto\e^{-(r-r_\mathrm{true})^2/2\sigma^2}$
with $\sigma=0.005$.

This analysis omits multipoles $24\leqslant\ell\leqslant29$, due to constraints from both
the low-$\ell$ and high-$\ell$ software. For the pixel-based approach at
low-$\ell$, the Healpix documentation recommends using
$\ell_\mathrm{max}=3N_\mathrm{side}-1=23$ to avoid oversampling each pixel,
while PolSpice introduces mode-coupling effects in the power spectrum for
$\ell\lesssim30$.

\begin{figure} 
    \begin{center} 
        \beginpgfgraphicnamed{f6} 
        \input{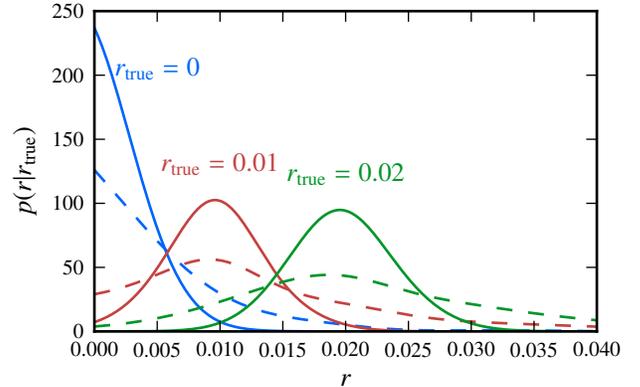}
    \endpgfgraphicnamed 
    \end{center}
    
    \caption{ We estimate the distributions for
    $r$ from Equation \ref{eq:bayes}, and use a Gaussian kernel density
    estimator to obtain a continuous curve, which returns a normalized
    probability density by default. The dashed lines are the
    distributions given only the low-$\ell$ data ($2\leqslant\ell\leqslant23$), while the
    solid curves are a product of the low-$\ell$ distributions and a Gaussian
    estimate of the high-$\ell$ ($30\leqslant\ell\leqslant100$) distribution with mean
    $\mu=r_\mathrm{true}$ and standard deviation $\sigma=0.005$. The solid curve
    for $r_\mathrm{true}=0$ gives a 95\% C.L.~upper limit $r<0.008$, while for
    the other solid curves the 95\% C.L.~limits are $0.006<r<0.014$ and
    $0.015<r<0.024$.} 

    \label{fig:prob_rtrue} 

\end{figure}

With the distribution for $r_\mathrm{true}=0$, we obtain the range of potential
false positive results, which suggests a definition of our upper limit as the
value of $r$ that contains 95\% of the area under the curve from $r=0$, yielding
$r<0.017$ for low-$\ell$ data alone, and $r<0.008$ using the high-$\ell$
simulations as well.  Similarly, we estimate the expected range of recovered $r$
for $r_\mathrm{true}=0.01$, and find that the 68\% confidence interval, defined
as the percentiles $(0.16, 0.5, 0.84)$, gives $r=0.011^{+0.011}_{-0.007}$
($r=0.010^{+0.004}_{-0.004}$ including high-$\ell$).  These distributions are
displayed in Figure \ref{fig:prob_rtrue}.  Using a simple likelihood ratio test
based on these curves with $\mathscr L(r)\equiv p(r|r_\mathrm{true}=0.01)$, we
find a likelihood ratio for the low-$\ell$ data with $\mathscr L(0)/\mathscr
L(0.01)=0.52$, corresponding to a PTE of $0.13$, or a significance of
$1.2\sigma$.  Including high-$\ell$ data gives a projected $\mathscr
L(0)/\mathscr L(0.01)=0.07$, corresponding to a PTE of $0.01$, or a significance
of $2.3\sigma$.

\section{Conclusions}

In this work, we have simulated CMB polarization maps on a cut sky contaminated
by diffuse foreground emission and Gaussian white noise.  We recovered the input
CMB by solving for maximum-likelihood maps using a pixel-based maximum
likelihood method tailored to measure the reionization peak ($\ell\leqslant23$).  We
have found the following.  

\begin{itemize} 
    
    \item We found unbiased estimates of
            the amplitudes of scalar and tensor fluctuations with a
            four-frequency maximum likelihood analysis accounting for CLASS-like
            noise, a cut sky, and foregrounds with spatially varying spectral
            indices.

    \item We have found that, by using only polarization the reionization peak
        ($2\leqslant\ell\leqslant23$), data from a five-year CLASS-like experiment would
        produce a 95\% upper limit of $r<0.017$, which improves to $r<0.008$
        when we include high-$\ell$ data ($30\leqslant\ell\leqslant100$).

    \item Our pixel-based low-multipole maximum likelihood solution is
        straightforward to run and implement, and is stable to foreground
        complications, i.e.~variable spectral indices and polarization
        fractions.

\end{itemize}

Although CLASS's primary purpose is constraining the amplitude of
tensor fluctuations, a foreground-cleaned measurement of CMB polarization on the
largest scales can help constrain several other cosmological parameters.
Our cosmic variance-limited measurement of the large-scale
E-mode power spectrum will improve on the current
\wmap{} and \planck{} measurements of
the optical depth $\tau$ to reionization by a factor of four. Not only can
optical depth measurements constrain cosmic reionization scenarios, but it can
also further constrain the amplitude of primordial density fluctuations $A_s$,
whose precision is limited by its degeneracy with $\tau$. With a
tightly constrained $A_s$ measurement, comparisons with the amplitude of large
scale structure of the Universe, $\sigma_8$, can be used to infer the sum of
neutrino masses to $0.015\ev$ \citep{mnu}. 
Finally, the observed anomalies in the
temperature fluctuations on large angular scales require polarization
measurements on the same scales to probe physics beyond standard {\lcdm}
\citep{wmapanomalies, planckisotropy}.

\acknowledgments{ We acknowledge the National Science Foundation Division of
    Astronomical Sciences for their support of CLASS under Grant Numbers 0959349
    and 1429236.  The CLASS project developed technology under several previous
    NASA grants, and NASA provides ongoing detector support for civil servants.
    Detector development work at JHU was funded by NASA grant number NNX14AB76A.
    CLASS is located in the Parque Astron\'omica Atacama in northern Chile under
    the auspices of the Comisi\'on Nacional de Investigaci\'on Cient\'\i fica y
    Tecnol\'ogica de Chile (CONICYT). D.J.W.~receives support from the Maryland
    Space Grant Consortium.  T.~E.-H.~receives support from a National Science
    Foundation Astronomy and Astrophysics Postdoctoral Fellowship.  
  }

\appendix

\section{Signal-to-Noise on Large Angular Scales} 

\label{sec:appendix_a} 

The purpose of this Appendix is to demonstrate that in the large-noise limit,
the signal-to-noise of B-mode polarization measurements increases with
decreasing multipole $\ell$.  Cosmic variance, the uncertainty in the observed
CMB anisotropy due to it being a single realization of a Gaussian random field,
is a limitation for large scale experiments in particular since the uncertainty
in the power spectrum scales as $\ell^{-1/2}$ in the cosmic variance limit.
For a beam window function $b_\ell$ and noise power spectrum $N_\ell$, the
signal-to-noise per multipole of a power spectrum is \citep{knox} 

\begin{equation} 
    \frac{C_\ell b_\ell^2} {(C_\ell b_\ell^2 + N_\ell)
        \sqrt{\frac2{f_\mathrm{sky}(2\ell+1)}}}
        =\frac{\sqrt{f_\mathrm{sky}(\ell+1/2)}}{1+N_\ell/(C_\ell b_\ell^2)}.
    \label{eq:s-to-n}
\end{equation}

For $C_\ell b_\ell^2\gg N_\ell$, i.e.~in the cosmic variance limit, the
$\ell^{1/2}$ factor dominates the signal-to-noise, but when the power spectra of
the noise and CMB are comparable the relationship is not as simple, and the
projected significance can sometimes increase at large angular scales. 

CMB polarization experiments are far from the cosmic variance measurement limits
of the B-mode power spectrum, so at large angular scales the signal-to-noise in
Equation \ref{eq:s-to-n} becomes 

\begin{equation} 
    \frac{C_\ell b_\ell^2\sqrt{f_\mathrm{sky}(\ell+1/2)}}{N_\ell}. 
\end{equation} 

For low multipoles the BB spectrum roughly scales with multipole as
$C_\ell\propto\ell^{-2}$, so that signal-to-noise scales as $\ell^{-3/2}$, in
contrast with the $\ell^{1/2}$ scaling found in the cosmic variance limit.

\section{Dust Spectral Index Variation} \label{sec:appendix_b}

\defcitealias{polfreq}{Planck Int XXII}

The purpose of this Appendix is to demonstrate that the observed variation in
dust spectral index observed by \planck{} \citep[][\citetalias{polfreq}
hereafter]{polfreq} is consistent with noise, and does not indicate any
intrinsic variation.  To estimate the spectral index of thermal dust emission,
the \planck{} team subdivided mid-latitude data into 400 overlapping patches of
sky \citepalias{polfreq}. They subtracted lower frequency data from
dust-dominated bands to remove the achromatic CMB component and then took ratios
of the residual at two different frequencies, $\nu_1$ and $\nu_2$, denoted
\begin{align} R_{\nu_0}(\nu_2,\nu_1)&=\frac{I_{\nu_2}-I_{\nu_0} }
    {I_{\nu_1}-I_{\nu_0} } \\
    &\simeq\left(\frac{\nu_2}{\nu_1}\right)^{\beta_\mathrm D}
    \frac{B_{\nu_2}(T_\mathrm D)}{B_{\nu_1}(T_\mathrm D)} \label{eq:color_ratio}
\end{align} where $\beta_\mathrm D$ is the spectral index of the dust,
$T_\mathrm D$ is the dust temperature, and $I_\nu$ is the total signal intensity
at frequency $\nu$.  To remove the CMB contribution from the dust-dominated
bands, they subtracted $I_{\nu_0}$ from $I_{\nu_2}$ and $I_{\nu_1}$ before
taking a ratio, and assumed the dust follows a modified blackbody spectrum
$\propto\nu^{\beta_\mathrm D}B_\nu(T_\mathrm D)$.  They estimated the dust
temperature using $R^\mathrm I(3000,857)$ with DIRBE data as the high-frequency
template, while they computed spectral indices using $R^\mathrm
{\{I,P\}}_{100}(353,217)$. Using $T_\mathrm D$, they implicitly solved for
$\beta_\mathrm D$ for each patch.

\begin{figure*} 
    \centering 
    \beginpgfgraphicnamed{f7} 
    \input{planck_figs.pgf}
    \endpgfgraphicnamed 
    \centering 
    \label{fig:beta_hists} 
    \caption{We reproduce
    Figure 8 of \citetalias{polfreq} on the left, and use the mean and standard
    deviation of $\beta_\mathrm{D}$ listed in the paper to reproduce the full
    histogram on the right, along with subsets corresponding to lower limits on
    signal fluctuation $\sigma^\mathrm P_{353}$. The data with lower limits
    $30\uk$ and $40\uk$ correspond to the same lower limits explored in \S B of
    \citetalias{polfreq}, and yield $\beta_\mathrm D$ fluctuations consistent
    with the $\beta_\mathrm D$ dispersion measured in Monte Carlo simulations
    with a \textit{constant} $\beta_\mathrm D$ and \planck{} noise.  }
\end{figure*}

The resulting distribution of $\beta_\mathrm D$ from \planck{} has a $1\sigma$
dispersion of $\sigma_{\beta_\mathrm D}=0.17$, which at face value suggests true
on-sky spectral index variation. However, each sky patch has an associated local
dispersion $\sigma^\mathrm P_{353}$ in the 353 GHz dust template brightness,
with higher values corresponding to higher signal-to-noise, and
\citetalias{polfreq} claim that low signal-to-noise values with
$\sigma_{353}^\mathrm P<20\uk$ are dominated by instrument noise.  Additionally,
in Appendix B of \citetalias{polfreq}, Monte Carlo simulations of dust
polarization with \textit{constant} $T_\mathrm D$ and $\beta_\mathrm D$ and
instrument noise give a  $1\sigma$ dispersion $\sigma_{\beta_\mathrm D}=0.07$ in
$\beta_\mathrm D$ (which is therefore entirely due to instrument noise) for sky
patches with signal $\sigma^\mathrm P_{353}>30\uk$.

Using data from the top plot of Figure 8 of \citetalias{polfreq} and the
relation from Equation \ref{eq:color_ratio} fixing $T_\mathrm D$ to be constant,
we rescale $R^\mathrm P_{100}(353,217)$, assuming a log-linear relation between
$R_{100}^\mathrm P(353,217)$ and $\beta_\mathrm{D}$.  We rescale
$\beta_\mathrm{D}$ to have the quoted mean and standard deviation of
$\beta_\mathrm{D}=1.59\pm0.17$.  We reproduce Figure 8 of \citetalias{polfreq}
in our Figure \ref{fig:beta_hists}. Using these estimates, we compute the
standard deviation of subsets of the data with $\sigma^\mathrm
P_{353}>[20,30,40]\uk$, obtaining $\sigma^{\mathrm
P}_{\beta_\mathrm{D}}=[0.12,0.07,0.05]$, respectively. We observe that the
observed variation of $\sigma^{\mathrm P}_{\beta_\mathrm D}=0.07$ for
$\sigma^\mathrm P_{353}>30\uk$ is entirely consistent with effects of instrument
noise seen in the \citetalias{polfreq} Monte Carlo simulations and, therefore,
with the hypothesis of a constant dust spectral index in the polarization data.
Based on this, we claim that our model of dust spatial spectral variation with
$\sigma^{\mathrm P}_{\beta_\mathrm D}=0.01$ as shown in Figure \ref{fig:regions}
is consistent with the latest \planck{} data.

\bibliographystyle{apj_hyperref}
\renewcommand{\eprint}[1]{\href{http://arxiv.org/abs/#1}{#1}}
\newcommand{\ISBN}[1]{\href{http://cosmologist.info/ISBN/#1}{ISBN: #1}}
\newcommand{\adsurl}[1]{\href{#1}{ADS}} \bibliography{fg}

\end{document}